# A search for $B_{s0}^*$ and $B_{s1}^*$ through the $K^-p$ interaction

**Min Yuan**[1,a], **Yin Huang**[1,b]

[1] School of Physical Science and Technology, Southwest Jiaotong University, Chengdu 610031, China

**Abstract** Studying heavy-quark hadrons is crucial due to the nonperturbative nature of low-energy QCD, with Heavy-Quark Symmetry (HQS) serving as a key framework for understanding their spin and flavor symmetries. However, a key issue is that the theoretically expected $B^{(*)}\bar{K}$ molecular states have not yet been observed, although they are considered the bottom-quark counterparts of the observed $\bar{D}^{(*)}\bar{K}$ molecular states (corresponding to $D_{s0}(2317/2460)^-$), which challenges the universality of HQS. The main goal of this work is to search for the theoretically predicted $B\bar{K}$ and $B^*\bar{K}$ molecular states, namely $B_{s0}^*(5725)$ and $B_{s1}^*(5778)$, via the reactions $K^-p \to \Lambda_b^0 B_{s0}^*$ and $K^-p \to \Lambda_b^0 B_{s1}^*$. Within an effective Lagrangian framework, we compute the relevant cross sections, considering $t$-channel $B^{(*)}$ exchanges and $K^-p$ initial-state interactions (ISI). The results show that the production cross sections of $B_{s0}^*(5725)$ and $B_{s1}^*(5778)$ can reach the order of 0.01 nb, and we suggest that experiments searching for $B_{s0}^*(5725)$ are best performed at $P_{K^-} = 12.18$ GeV, while higher energies are most favorable for producing $B_{s1}^*(5778)$. The ISI play a crucial role, as they not only significantly enhance the production cross sections of $B_{s0}^*(5725)$ and $B_{s1}^*(5778)$ (by roughly one order of magnitude) but also markedly affect the angular distributions of the produced particles. We also calculated the production cross sections of the conventional quark–antiquark states $B_{s0}^*(5700)$ and $B_{s1}^*(5720)$, which are found to be nearly the same as those of $B_{s0}^*(5725)$ and $B_{s1}^*(5778)$. Although their internal structures remain ambiguous, these results can inform future experimental searches at CERN and J-PARC.

## 1 Introduction

In recent decades, the continuous advancement of experimental techniques has led to the discovery of a large number of exotic hadronic states [1]. These so-called exotic hadrons exhibit more intricate internal structures than conventional quark configurations, which describe mesons as quark–antiquark pairs and baryons as three-quark combinations. As a result, understanding their internal structure has emerged as a central issue in contemporary particle physics. Due to quark confinement, quarks cannot be observed directly. More importantly, the strong coupling constant becomes large at low energies, rendering perturbative techniques inapplicable. Consequently, traditional perturbation theory fails to provide accurate descriptions of quark interactions, necessitating the use of various non-perturbative approaches, such as lattice quantum chromodynamics and effective field theories. Among these, heavy-quark symmetry (HQS) [2] has proven especially useful for studying hadrons containing heavy quarks.

HQS is an approximate symmetry of quantum chromodynamics (QCD) that emerges in the limit where the heavy quark mass greatly exceeds the QCD scale, $\Lambda_{\rm QCD} \sim 200$ MeV. In this limit, the spin and flavor of the heavy quark become conserved under the strong interaction, significantly simplifying the analysis of hadronic systems with heavy quarks. Specifically, in the limit $m_Q \to \infty$, the heavy quark's mass plays a negligible role in strong dynamics, and hadronic systems become insensitive to the heavy quark's flavor (e.g., $c$ or $b$). This symmetry leads to analogous internal structures between hadrons containing a $c$ quark (e.g., $D$, $\Lambda_c$) and those with a $b$ quark (e.g., $B$, $\Lambda_b$), allowing properties of $b$-quark hadrons to be inferred from their $c$-quark counterparts.

Nevertheless, HQS faces notable limitations when confronted with certain experimental observations, which chal-lenges its broad applicability as a foundational framework for describing strong interactions. For instance, heavy quark flavor symmetry (HQFS) predicts the existence of a family of hidden-bottom pentaquark molecular states based on the discovery of several hidden-charm candidates, including $P_c(4312)$, $P_c(4380)$, $P_c(4440)$, $P_c(4457)$, $P_{cs}(4338)$, and $P_{cs}(4459)$ [3–19]. However, no experimental evidence has been reported for their hidden-bottom counterparts. Additionally, heavy quark spin symmetry (HQSS) predicts a partner state of the $X(3872)$ with quantum numbers $J^{PC} = 2^{++}$, primarily composed of a $D^*\bar{D}^*$ molecular configuration [20]. Although the Belle Collaboration reported a possible signal in the $\gamma\gamma \to \gamma\psi(2S)$ channel in 2022, its statistical significance was only $2.8\sigma$ [21], and further experimental confirmation is required.

[a] e-mail: yuanm@my.swjtu.edu.cn
[b] e-mail: huangy2019@swjtu.edu.cn (corresponding author)

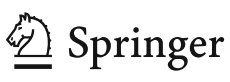

Another important question pertains to the existence of $B\bar{K}$ ($B^*_{s0}$ with $J^P = 0^+$) and $B^*\bar{K}$ ($B^*_{s1}$ with $J^P = 1^+$) hadronic molecules. This issue gained prominence following the discovery of their HQFS counterparts, the $D^*_{s0}(2317)^-$ ($J^P = 0^+$) and $D^*_{s1}(2460)^-$ ($J^P = 1^+$) [22–26]. Due to their unconventional properties, these charm-strange mesons have been widely interpreted as $S$-wave $\bar{D}\bar{K}$ and $\bar{D}^*\bar{K}$ molecular states, respectively [27–33].

Theoretical studies indicate that the $B\bar{K}$ and $B^*\bar{K}$ molecules, like their charm analogs, are expected to lie about 40–60 MeV below the corresponding thresholds and primarily decay via isospin-violating $\pi B_s^{(*)}$ channels. Using a heavy-quark chiral Lagrangian and non-perturbative unitarized coupled-channel scattering amplitudes, poles were extracted on the appropriate Riemann sheets in Ref. [28], predicting a $B\bar{K}$ molecule with a mass of $5.725 \pm 0.039$ GeV. A $B^*\bar{K}$ molecule with mass $5.778 \pm 0.007$ GeV was later predicted using the same framework [29]. The decays into $\pi B_s^{(*)}$ were also examined using the Bethe–Salpeter approach [34] and compositeness conditions [35,36]. Furthermore, assuming these are $B^{(*)}\bar{K}$ molecules, their mass spectra and decay properties have been systematically analyzed in low-energy chiral effective theory [37–39]. Another study [40] also supported their subthreshold nature and narrow widths, consistent with a molecular interpretation.

Conversely, some studies argue that the absence of experimental evidence for molecular states dominated by $B\bar{K}$ and $B^*\bar{K}$ components – namely $B^*_{s0}$ and $B^*_{s1}$ – suggests that such configurations may not exist. Instead, these states are interpreted as conventional $b\bar{s}$ mesons. In the framework of QCD sum rules, their masses have been estimated to be approximately $M_{B^*_{s0}} = 5700$ MeV and $M_{B^*_{s1}} = 5720$ MeV [41], with corresponding coupling strengths $g_{B^*_{s0}B\bar{K}} = 20.0 \pm 7.4$ GeV and $g_{B^*_{s1}B^*\bar{K}} = 18.1 \pm 6.1$ GeV [42]. Nonetheless, no experimental confirmation of these conventional bottom-strange mesons has been reported to date [1].

In light of this ambiguity, a central question emerges: do the $B^{(*)}\bar{K}$ molecular states associated with $B^*_{s0}$ and $B^*_{s1}$ truly exist, or should they be regarded as conventional quark–antiquark mesons? Resolving this question is essential not only for clarifying the nature of these states, but also for testing the broader implications of heavy quark symmetry (HQS). Therefore, direct experimental investigation of their production mechanisms becomes critically important. High-energy kaon beams at facilities such as OKA@U-70 [43], SPS@CERN [44], and the recently launched AMBER@CERN [45], as well as potential upgrades at J-PARC, offer the necessary energy range to produce $B^*_{s0}$ and $B^*_{s1}$. This makes searches for these states in the reactions $K^-p \to \Lambda_b^0 B^*_{s0}$ and $K^-p \to \Lambda_b^0 B^*_{s1}$ experimentally feasible. If HQS holds, our study provides a concrete pathway for guiding experimental identification of these states. To enhance the reliability of such predictions, it is crucial to account for the initial-state interaction (ISI) of the $K^-p$ system, given the abundance of experimental data on $K^-p$ elastic scattering in the relevant energy region.

This paper is organized as follows: in Sect. 2, we present the theoretical formalism. In Sect. 3, the numerical result and discussions are given, followed by conclusions in the last section.

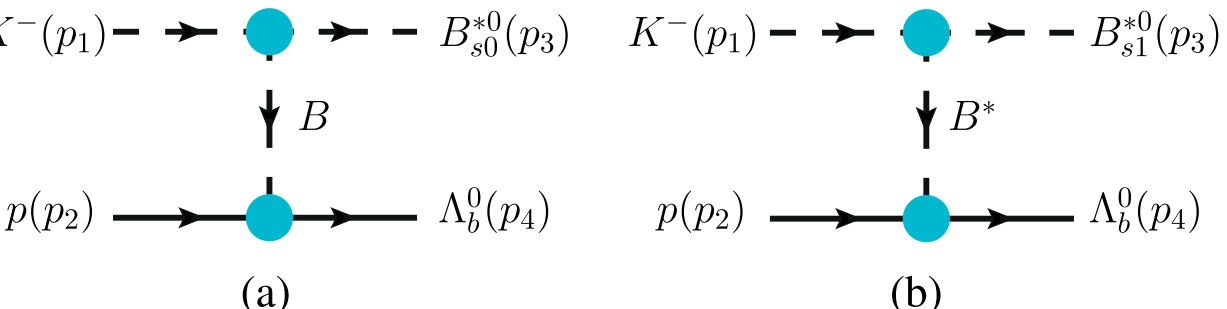

**Fig. 1** The Feynman diagrams illustrate the production mechanisms of $B^*_{s0}$ and $B^*_{s1}$ in the $K^-p \to \Lambda_b^0 B^*_{s0}$ (left) and $K^-p \to \Lambda_b^0 B^*_{s1}$ (right) reactions. The definitions of the kinematics ($p_1$, $p_2$, $p_3$, $p_4$) used in the calculations are also provided

## 2 Theoretical formalism

In Fig. 1, we present the Feynman diagrams for the production of $B^*_{s0}$ and $B^*_{s1}$ through the reactions $K^-p \to \Lambda_b^0 B^*_{s0}$ (left) and $K^-p \to \Lambda_b^0 B^*_{s1}$ (right), respectively, focusing on the exchange of $B$ and $B^*$ mesons in the $t$-channel. The contributions from the $s$-channel and $u$-channel interactions are neglected, primarily because these processes would involve the production of multi-quark baryons containing a $b\bar{b}$ pair, corresponding to $P_{bs}$ baryon states with a minimum mass of 11.35 GeV. To date, no experimental or theoretical information is available for such states. Including it would introduce extremely large uncertainties, rendering quantitative predictions essentially meaningless. Consequently, the analysis focuses on the exchange effects of $B$ and $B^*$ mesons in the $t$-channel.

To evaluate the amplitude of the Feynman diagram shown in Fig. 1, we begin by constructing the effective Lagrangian densities for the relevant interaction vertices. For the $\Lambda_b pB$ and $\Lambda_b pB^*$ couplings, we adopt the following effective Lagrangians [46]:

$$\mathscr{L}_{\Lambda_b pB} = ig_{\Lambda_b pB}\bar{\Lambda}_b\gamma_5 pB^0 + \text{H.c.}, \tag{1}$$

$$\mathscr{L}_{\Lambda_b pB^*} = g_{\Lambda_b pB^*}\bar{\Lambda}_b\gamma^\mu pB^{*0}_\mu + \text{H.c.}, \tag{2}$$

where the coupling constants are taken as $g_{\Lambda_b pB}$= − 13.41 and $g_{\Lambda_b pB^*}$= − 5.63. These values are derived from an SU(4)-invariant Lagrangian [47], using the empirical couplings $g_{\pi NN} = 13.5$ and $g_{\rho NN} = 3.25$ as input.

Besides, it is essential to construct the effective Lagrangians for the $\bar{K}BB^*_{s0}$ and $\bar{K}B^*B^*_{s1}$ vertices. Since the $B^*_{s0}$ and $B^*_{s1}$ are interpreted as $S$-wave hadronic molecular states of $B\bar{K}$ and $B^*\bar{K}$, respectively [28,29,34–40], the corresponding effective interaction Lagrangians, as shown in Fig. 1, take the form:

$$\mathscr{L}_{\bar{K}BB^*_{s0}} = g_{\bar{K}BB^*_{s0}}\bar{K}BB^*_{s0}, \tag{3}$$

$$\mathscr{L}_{\bar{K}B^*B^*_{s1}} = g_{\bar{K}B^*B^*_{s1}}\bar{K}B^{*\mu}B^*_{s1,\mu}. \tag{4}$$

The coupling constants $g_{\bar{K}BB^*_{s0}}$=23.442 GeV and $g_{\bar{K}B^*B^*_{s1}}$= 23.572 GeV are adopted from Refs. [28,29], in which the $B^*_{s0}$ and $B^*_{s1}$ are dynamically generated from the $B\bar{K}$ and $B^*\bar{K}$–$\eta B_s$ interactions within the heavy chiral unitary framework. For comparison, the corresponding coupling constants for the scenario where $B^*_{s0}$ and $B^*_{s1}$ are treated as conventional $b\bar{s}$ states, as determined by QCD sum rules [41,42], are also listed in Table 1. These values will be used in Sect. 3 to calculate and compare the production cross sections under both interpretations.

Importantly, these studies utilize the same formalism that successfully describes the $D^*_{s0}(2317)$ and $D^*_{s1}(2460)$ as dynamically generated from $KD$ and $KD^*$–$\eta D_s$ interactions. The framework incorporates heavy-quark flavor symmetry (HQFS) to relate the charm and bottom sectors. In this approach, the $B^*_{s0}$ and $B^*_{s1}$ states naturally emerge as HQFS partners of the experimentally observed $D^*_{s0}(2317)$ and $D^*_{s1}(2460)$, respectively. The model parameters are fixed by reproducing the charm-sector masses and then applied to the bottom sector, leading to predictions for the bottom-strange states.

Consequently, the predicted masses of the bottom partners are $M_{B^*_{s0}} = 5.725 \pm 0.039$ GeV and $M_{B^*_{s1}} = 5.778 \pm 0.007$ GeV. Here and throughout this work, we refer to these states as $B^*_{s0}(5725)$ and $B^*_{s1}(5778)$, respectively. Based on this, we investigate their production in the reactions $K^-p \to \Lambda^0_b B^*_{s0}(5725)$ and $K^-p \to \Lambda^0_b B^*_{s1}(5778)$. Once observed, such states will serve as a crucial test for heavy-quark symmetry.

In the $K^-p \to \Lambda^0_b B^*_{s0}(5725)$ and $K^-p \to \Lambda^0_b B^*_{s1}(5778)$ reactions, we need to include form factors because hadrons are not pointlike particles. For the exchanged $B$ and $B^*$ mesons, we apply a widely used monopole form factor, which is written as [48,49]:

$$F_i(q_i^2) = \frac{\Lambda_i^2 - m_i^2}{\Lambda_i^2 - q_i^2}, \quad i = B, B^*, \tag{5}$$

where $q_i$ and $m_i$ are the four-momentum and the mass of the exchanged $B^{(*)}$ meson, respectively. The cutoff $\Lambda_i$ characterizes the finite size of the hadrons, and is empirically a few hundred MeV larger than the mass of the exchanged meson. Hence, we adopt

$$\Lambda_i = m_i + \alpha\Lambda_{\text{QCD}}, \tag{6}$$

with the QCD energy scale $\Lambda_{\text{QCD}} = 220$ MeV. The dimensionless parameter $\alpha$ reflects the nonperturbative nature of QCD at low energies and can only be determined from experimental data. In this work, $\alpha$ is treated as a free parameter and will be discussed later.

The propagator for the $B$ meson is given by

$$G_B(q) = \frac{i}{q^2 - m_B^2}, \tag{7}$$

and for the $B^*$ meson exchange, we use the propagator

$$G^{\mu\nu}_{B^*}(q) = \frac{i(-g^{\mu\nu} + q^\mu q^\nu/m^2_{B^*})}{q^2 - m^2_{B^*}}, \tag{8}$$

where $\mu$ and $\nu$ are the Lorentz indices of the $B^*$ meson.

With all these ingredients, the invariant scattering amplitude of the $K^-p \to \Lambda^0_b B^*_{s0}(5725)$ and $K^-p \to \Lambda^0_b B^*_{s1}(5778)$ reactions demonstrated in Fig. 1 can be constructed as

$$\begin{aligned}\mathscr{M}_{B^*_{s0}} &= -g_{\Lambda_b pB}\bar{u}(p_4, s_{\Lambda^0_b})\gamma_5 u(p_2, s_p)\frac{1}{(p_3 - p_1)^2 - m_B^2}\\ &\quad\times g_{\bar{K}B^*_{s0}B}F_B^2\left((p_3 - p_1)^2\right),\end{aligned} \tag{9}$$

$$\begin{aligned}\mathscr{M}_{B^*_{s1}} &= ig_{\Lambda_b pB^*}\bar{u}(p_4, s_{\Lambda^0_b})\gamma_\mu u(p_2, s_p)\frac{1}{(p_3 - p_1)^2 - m^2_{B^*}}\\ &\quad\times\left(-g^{\mu\nu} + \frac{(p_3 - p_1)^\mu(p_3 - p_1)^\nu}{m^2_{B^*}}\right)g_{\bar{K}B^*_{s1}B^*}\\ &\quad\times F^2_{B^*}\left((p_3 - p_1)^2\right)\epsilon^*_\nu(p_3, s_{B^*_{s1}}),\end{aligned} \tag{10}$$

where $u$ and $\epsilon$ are the Dirac spinor and polarization vector, respectively. $s_{\Lambda^0_b}$, $s_{B^*_{s1}}$, and $s_p$ are the spins of the outgoing $\Lambda^0_b$, $B^*_{s1}$ and the initial proton, respectively.

To enhance the reliability of our predictions, we incorporate corrections to the Born amplitudes in Eqs. (9) and (10), which arise from the $K^-p \to K^-p$ interactions in the initial-state (called initial-state interactions - ISI). These interactions modify the scattering amplitude through both elastic and inelastic contributions. As illustrated in Fig. 2, the black square represents the full $K^- p \to K^- p$ interaction, including the tree-level elastic part and inelastic effects (denoted by the ellipses). Such corrections are crucial for a realistic description of the production process.

**Table 1** Key parameters for the hadronic molecular and conventional quark–antiquark interpretations of the $B^*_{s0}$ and $B^*_{s1}$ states

| Parameter | Molecule state [28,29] | $b\bar{s}$ State [41,42] |
|---|---|---|
| Masses | | |
| $M_{B^*_{s0}}$ (GeV) | $5.725 \pm 0.039$ | 5.700 |
| $M_{B^*_{s1}}$ (GeV) | $5.778 \pm 0.007$ | 5.720 |
| Coupling constants | | |
| $g_{B^*_{s0}B\bar{K}}$ (GeV) | 23.442 | $20.0 \pm 7.4$ |
| $g_{B^*_{s1}B^*\bar{K}}$ (GeV) | 23.572 | $18.1 \pm 6.1$ |

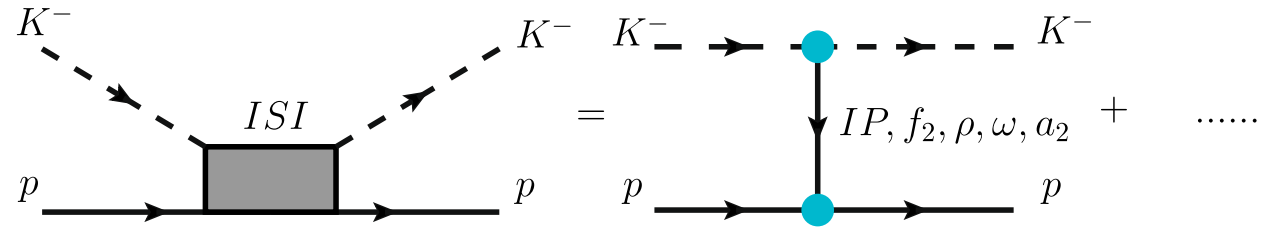

**Fig. 2** The Feynman diagram illustrates the initial-state interaction (ISI) mechanism of the $K^- p \to K^- p$ reaction (the part on the left-hand side of the equality sign). The right-hand side of the equality sign represents the tree-level scattering process, which includes the elastic and inelastic $K^- p$ interactions (indicated by the ellipses)

In this work, we follow a simplified approach proposed in Ref. [50], where the $K^- p \to K^- p$ elastic amplitude is modeled by considering only Pomeron and Reggeon exchanges. Despite its simplicity, this mechanism has proven sufficient to reproduce the elastic scattering data with high accuracy (see Fig. 1 and Fig. 2 in Ref. [50]). Based on the elastic amplitude thus obtained, the total cross section can be related to the forward scattering amplitude via the optical theorem, allowing for a consistent and accurate description of both elastic and total cross sections. Therefore, we simply calculate the elastic interaction $K^- p \to K^- p$ to represent the initial-state interaction of $K^- p \to K^- p$ in this work.

The underlying exchange mechanisms are depicted in the tree-level diagrams on the right-hand side of Fig. 2, where only contributions from the Pomeron and the $f_2, a_2, \rho$, and $\omega$ Reggeons are included. The full $K^- N \to K^- N$ amplitude can be written as the sum of these exchanges [50]:

$$T_{K^- N \to K^- N}(s,t) = A_{\mathrm{IP}}(s,t) + A_{f_2}(s,t) \pm A_{a_2}(s,t) + A_{\omega}(s,t) \pm A_{\rho}(s,t), \tag{11}$$

where $s$ is the squared center-of-mass energy and $t$ is the squared four-momentum transfer between the incoming and outgoing $K^-$ mesons. The upper (lower) signs apply to $K^- p$ ($K^- n$) scattering, respectively.

At high energies $\sqrt{s}$, each individual contribution to the $\bar{K}N \to \bar{K}N$ amplitude is parameterized as [50]:

$$A_i(s,t) = \eta_i \, s \, C_i^{\bar{K}N} \left(\frac{s}{s_0}\right)^{\alpha_i(t)-1} \exp\left(\frac{1}{2} B_i^{\bar{K}N} t\right), \tag{12}$$

where $i = \mathrm{IP}, f_2, a_2, \omega, \rho$ labels the exchanged trajectories. Here, $s_0 = 1$ GeV$^2$ is the energy scale, $\alpha_i(t) = \alpha_i(0) + \alpha_i' t$ is the linear Regge trajectory, and $\eta_i$ is the signature factor. The parameters $C_i^{\bar{K}N}$, $\alpha_i(0)$, $\alpha_i'$, $\eta_i$, and $B_i^{\bar{K}N}$ adopted from Ref. [50] are listed in Table 2, and are fitted to reproduce the elastic and total cross section data.

After taking into account the initial-state interaction in the $K^- p \to K^- p$ channel, the full amplitude can be expressed in the following form [50]:

$$\mathcal{M}_{\mathrm{full}} = \mathcal{M}_{B^*_{s0/s1}} + \frac{i}{16\pi^2 s} \int d^2 \vec{k}_t T_{K^- p \to K^- p}(s, k_t^2) \times \mathcal{M}_{B^*_{s0/s1}}(s, k_t^2), \tag{13}$$

where $\vec{k}_t$ represents the momentum transfer in the $K^- p \to K^- p$ reaction.

With the above ingredients, we proceed to compute the differential cross sections for the reactions $K^- p \to \Lambda_b^0 B^*_{s0}(5725)$ and $K^- p \to \Lambda_b^0 B^*_{s1}(5778)$ in the center-of-mass (CM) frame. The differential cross section is given by

$$\frac{d\sigma}{d\cos\theta} = \frac{m_p m_{\Lambda_b^0}}{8\pi s} \frac{|\vec{p}_{3,\mathrm{cm}}|}{|\vec{p}_{1,\mathrm{cm}}|} \left( \frac{1}{2} \sum_{s_p, s_{\Lambda_b^0}} |\mathcal{M}_{\mathrm{full}}|^2 \right), \tag{14}$$

where $\theta$ denotes the scattering angle of the outgoing meson ($B^*_{s0}(5725)$ or $B^*_{s1}(5778)$) with respect to the beam direction in the CM frame. The magnitudes of the three-momenta of the initial-state $K^-$ and final-state $B^*_{s0}(5725)$ or $B^*_{s1}(5778)$ in the CM frame are given by

$$|\vec{p}_{1,\mathrm{cm}}| = \frac{\lambda^{1/2}(s, m_{K^-}^2, m_p^2)}{2\sqrt{s}}, \tag{15}$$

**Table 2** The parameters of the Pomeron and Reggeon exchanges were determined based on elastic and total cross section data in Ref. [50]

| $i$ | $\eta_i$ | $\alpha_i(t)$ | $C_i^{\bar{K}N}$ (mb) | $B_i^{\bar{K}N}$ ($\mathrm{GeV}^{-2}$) |
|---|---|---|---|---|
| IP | i | $1.081 + (0.25\,\mathrm{GeV}^{-2})$t | 11.82 | 5.5 |
| $f_2$ | $-0.861 + i$ | $0.548 + (0.93\,\mathrm{GeV}^{-2})$t | 15.67 | 4.0 |
| $\rho$ | $-1.162 - i$ | $0.548 + (0.93\,\mathrm{GeV}^{-2})$t | 2.05 | 4.0 |
| $\omega$ | $-1.162 - i$ | $0.548 + (0.93\,\mathrm{GeV}^{-2})$t | 7.055 | 4.0 |
| $a_2$ | $-0.861 + i$ | $0.548 + (0.93\,\mathrm{GeV}^{-2})$t | 1.585 | 4.0 |

$$|\vec{p}_{3,\mathrm{cm}}| = \frac{\lambda^{1/2}(s, m^2_{B^*_{s0/s1}}, m^2_{\Lambda^0_b})}{2\sqrt{s}}, \tag{16}$$

where $\lambda(x, y, z) = (x - y - z)^2 - 4yz$ is the Källén function. The masses of the relevant particles used in this study are taken as $m_{K^-} = 493.68\,\mathrm{MeV}$, $m_p = 938.27\,\mathrm{MeV}$, and $m_{\Lambda^0_b} = 5619.60\,\mathrm{MeV}$.

## 3 Results and discussions

Since the $D^*_{s0}(2317)$ and $D^*_{s1}(2460)$ states have been observed in experiments [1], and are commonly interpreted as $DK$ and $D^*K$ hadronic molecules [27–33], there has been strong interest in identifying their heavy-quark flavor symmetry partners, namely the $B^*_{s0}$ and $B^*_{s1}$ states. Motivated by their predicted masses and the availability of current $K^-$ beam facilities, we propose to search for evidence of these states via the reactions $K^-p \to \Lambda^0_b B^*_{s0}(5725)$ and $K^-p \to \Lambda^0_b B^*_{s1}(5778)$.

At present, the primary source of uncertainty in our results is the undetermined parameter $\alpha$. Since $\alpha$ cannot be calculated from first principles, its value is typically obtained by fitting to experimental data. However, the two processes considered in this work have not been studied experimentally, making it difficult to determine $\alpha$ from their outcomes. It is worth noting that, as $\alpha$ characterizes the form factors of the exchanged $B$ and $B^*$ mesons, it can instead be extracted from other reactions involving the same exchanges. More importantly, the values of $\alpha$ obtained from such fits should ideally not induce a strong dependence when applied to other processes. Encouragingly, previous studies of the decay $X_b \to \gamma\Upsilon(1S, 2S, 3S)$, considering only $B$ and $B^*$ exchanges, show that the decay widths remain at the keV scale for $\alpha$ in the range 2–3 GeV [51], indicating a weak dependence on $\alpha$. This weak dependence is further confirmed in the study of the reactions $\Upsilon(5S, 6S) \to \gamma X_b$ [52], again considering only $B$ and $B^*$ exchanges. Using the same range, $\alpha = 2$–3, one can also reproduce the experimental decay width of $Z_b \to \Upsilon(2S)\pi$, corresponding to $(3.62^{+0.76}_{-0.59})\%$ of the total width [53], once more considering only $B$ and $B^*$ meson exchanges. Furthermore, by fitting the experimental data of the decays $Z_b/Z^{'}_b \to \Upsilon\pi$ and $h_b\pi$, the parameter $\alpha$, which is related solely to the form factors of the $B$ and $B^*$ mesons, can be constrained to the range 1.36–2.90 GeV, with central values determined to be 1.38 GeV and 2.90 GeV, respectively [54]. Motivated by these considerations, we perform calculations of the cross sections within the range $\alpha = 2.0$–2.9, and the results are presented in Fig. 3.

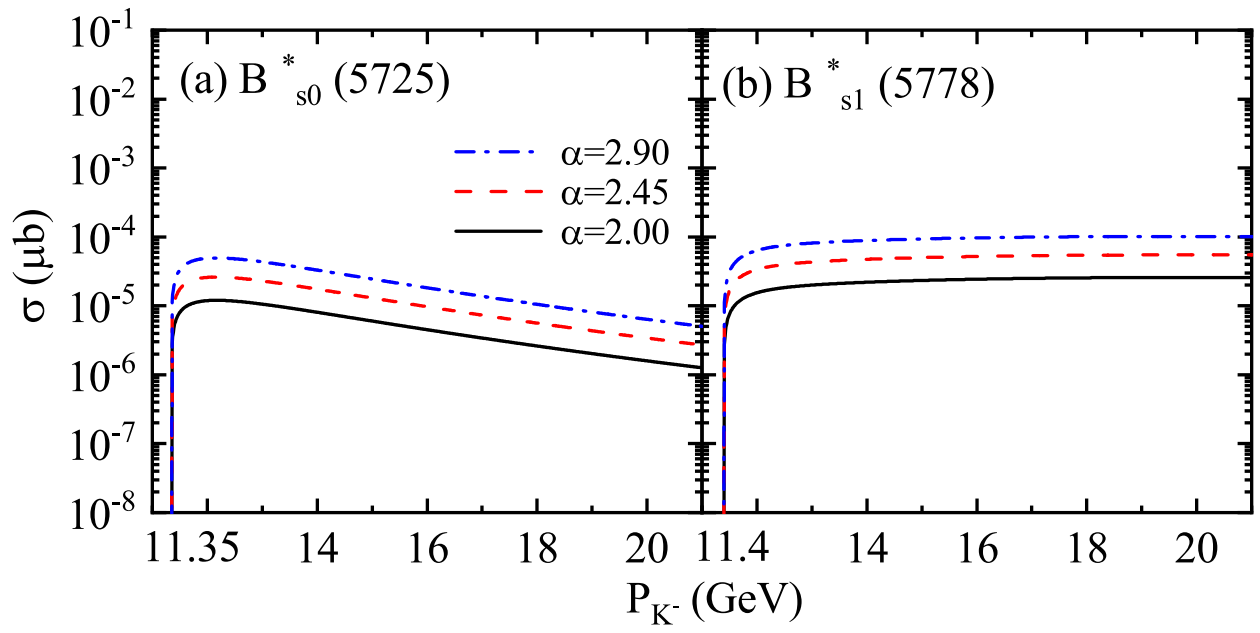

**Fig. 3** The total cross sections for the processes **a** $K^-p \to \Lambda^0_b B^*_{s0}(5725)$ and **b** $K^-p \to \Lambda^0_b B^*_{s1}(5778)$ are evaluated for $\alpha$ = 2.00, 2.45, and 2.90

In Fig. 3, we show three curves for representative values of $\alpha$ in the considered range, namely $\alpha = 2.00$, 2.45, and 2.90. The results indicate that, in the $K^-p \to \Lambda^0_b B^*_{s0}(5725)$ and $K^-p \to \Lambda^0_b B^*_{s1}(5778)$ reactions, the total production cross sections for $B^*_{s0}(5725)$ and $B^*_{s1}(5778)$ exhibit a pronounced and rapid rise near 11.35 GeV and 11.40 GeV, respectively. This sharp enhancement corresponds to the production thresholds of the two particles, signaling the opening of the available phase space. Since the mass of the $B^*_{s1}(5778)$ is larger than that of the $B^*_{s0}(5725)$, its production requires a higher threshold energy. As the energy increases further, the variation of the production cross sections gradually slows down due to the combined effects of phase-space saturation, propagator suppression, and form-factor damping at larger momentum transfer. We can also observe that the calculated cross sections exhibit a weak dependence on the parameter $\alpha$, as their production cross sections remain within the same order of magnitude. For example, at an incident $K^-$ beam energy of 14.0 GeV, the production cross section of $B^*_{s0}(5725)$ increases from 0.00806 nb for $\alpha = 2.0$ to 0.0329 nb for $\alpha = 2.9$, while that of $B_{s1}^*(5778)$ changes from 0.0219 to 0.0888 nb over the same range of $\alpha$ values.

Moreover, the results reveal that the line-shape behaviors of the two reactions are markedly different. When the $K^-$ beam momentum $P_{K^-}$ exceeds 12.18 GeV for $\alpha = 2.0$ (and 12.16 GeV for both $\alpha = 2.45$ and 2.90), the production cross section for $K^- p \to \Lambda_b^0 B_{s0}^*(5725)$ begins to decrease with increasing incident $K^-$ beam energy. In contrast, the cross section for $K^- p \to \Lambda_b^0 B_{s1}^*(5778)$ continues to increase slowly. This behavior indicates that $P_{K^-} \simeq 12.18$ GeV (or 12.16 GeV, depending on $\alpha$) corresponds to the optimal beam momentum for observing $B_{s0}^*(5725)$ in the $K^- p \to \Lambda_b^0 B_{s0}^*(5725)$ reaction. On the other hand, higher beam energies are more favorable for the production of $B_{s1}^*(5778)$, as its cross section keeps rising over the entire energy range considered.

The pronounced difference in the energy dependence of the cross sections can be traced back to the distinct production mechanisms of $B_{s0}^*(5725)$ and $B_{s1}^*(5778)$, both of which proceed via $t$-channel meson exchange, as illustrated in Fig. 1. Specifically, the production of $B_{s0}^*(5725)$ is mediated by $B$-meson exchange, while $B_{s1}^*(5778)$ is produced through $B^*$-meson exchange. Although both $B$ and $B^*$ are heavy-flavor mesons, they differ in spin and parity – $B$ being a scalar (spin-0) and $B^*$ a vector (spin-1) – which leads to fundamentally different coupling structures at the interaction vertices. As shown in Eqs. (3) and (4), the scalar vertex associated with $B$ exchange typically results in simple amplitudes with mild energy dependence, whereas the vector vertex from $B^*$ exchange involves Lorentz structures (here refer to $B^{*\mu} B_{s1,\mu}^*$). These structural differences yield distinct spin-summed amplitudes, ultimately leading to the observed variation in cross section behavior for the two processes.

We now turn to the effects of the initial-state $K^- p$ interaction on the results. To illustrate the influence of the $K^- p$ initial-state interaction (ISI), we compare the cross sections calculated with and without ISI for the cutoff $\alpha = 2.90$ in Fig. 4, for the $K^- p \to \Lambda_b^0 B_{s0}^*(5725)$ (Fig. 4a) and $K^- p \to \Lambda_b^0 B_{s1}^*(5778)$ (Fig. 4b) reactions. In both panels of Fig. 4, the solid black lines represent the pure Born contributions, while the dashed red lines show the full results including ISI.

The results show that the inclusion of $K^- p$ ISI leads to an enhancement of the cross sections for both reactions by approximately one order of magnitude. For a more detailed analysis, we examine the total cross section behavior at an incident $K^-$ meson energy of 14.0 GeV. The production cross section of $B_{s0}^*(5725)$ in the $K^- p \to \Lambda_b^0 B_{s0}^*(5725)$ reaction is calculated to be 0.0329 nb when ISI are accounted for, compared to 0.00333 nb when ISI effects are ignored-indicating an enhancement by a factor of 9.879. Likewise, the production cross section of $B_{s1}^*(5778)$ in $K^- p \to \Lambda_b^0 B_{s1}^*(5778)$ increases from 0.00898 nb (without ISI) to 0.0888 nb (with ISI), a similar factor difference of 9.888. A possible explanation for this significant enhancement is that the initial-state interaction (ISI) distorts the incoming wavefunction, thereby increasing its overlap with the final-state configuration and enhancing the transition probability [55,56]. These results indicate that the $K^- p$ initial-state interaction plays a non-negligible role in the production dynamics and is of great importance for understanding the search for other new particles via $K^- p$ interactions.

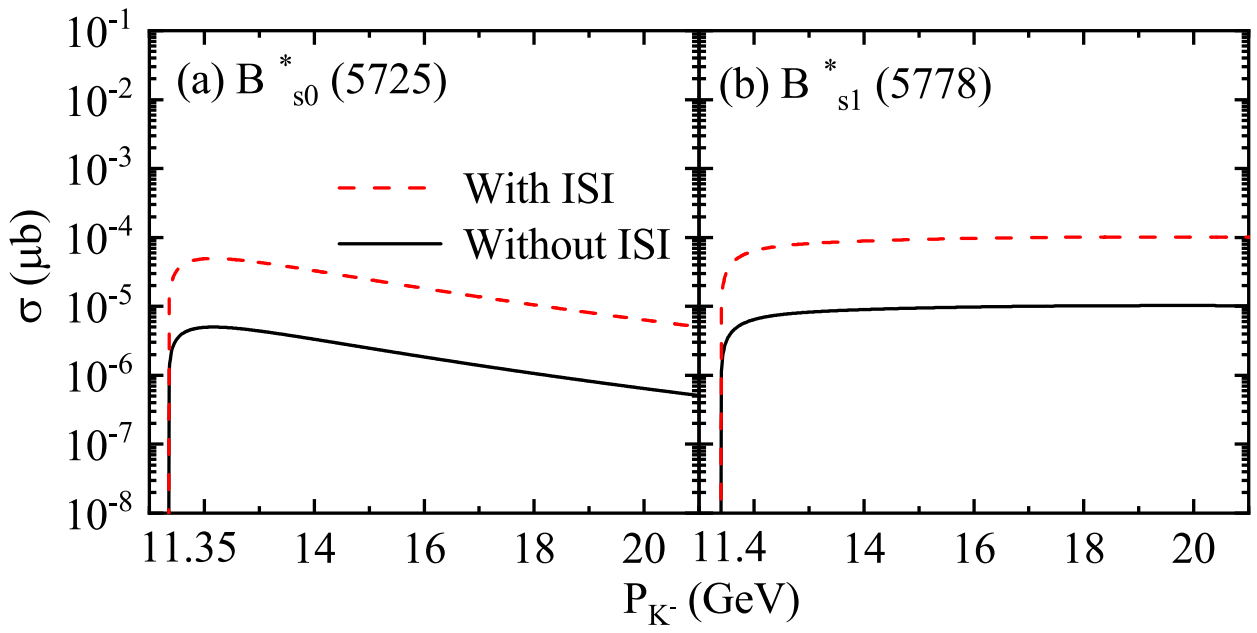

**Fig. 4** Cross sections with and without ISI for the processes **a** $K^- p \to \Lambda_b^0 B_{s0}^*(5725)$ and **b** $K^- p \to \Lambda_b^0 B_{s1}^*(5778)$ is evaluated for $\alpha = 2.90$

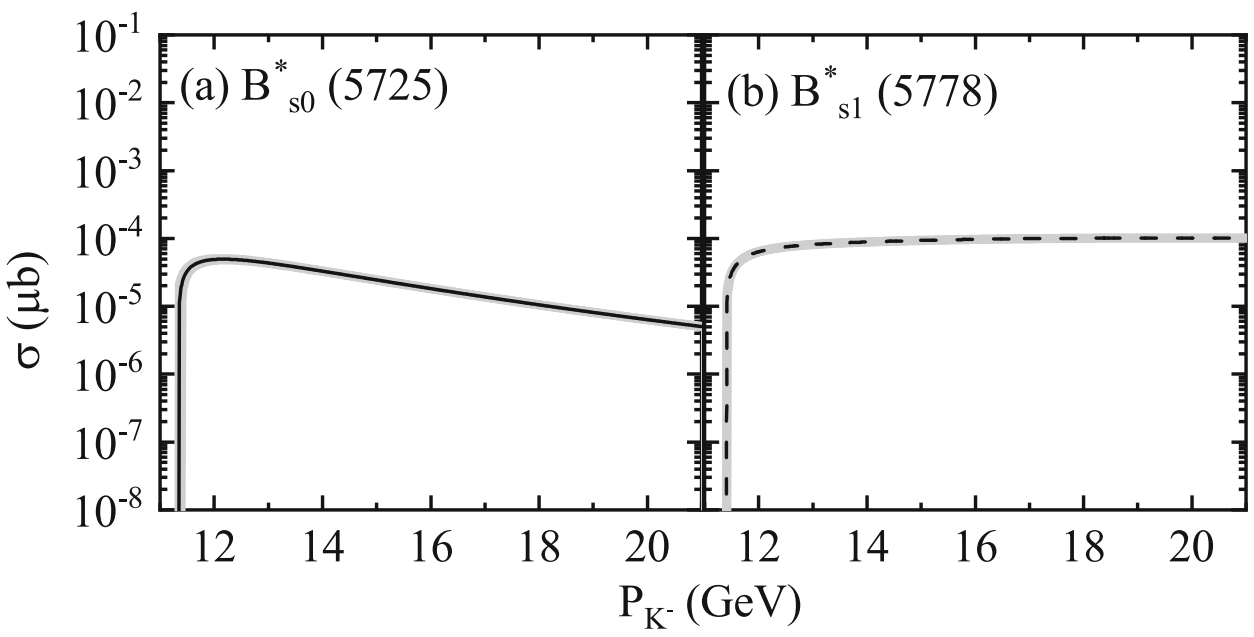

**Fig. 5** The total cross sections for the processes **a** $K^- p \to \Lambda_b^0 B_{s0}^*(5725)$ and **b** $K^- p \to \Lambda_b^0 B_{s1}^*(5778)$ are evaluated for $\alpha = 2.90$. The gray bands represent the uncertainty in the cross sections arising from the theoretical mass uncertainties $M_{B_{s0}^*} = 5.725 \pm 0.039$ GeV and $M_{B_{s1}^*} = 5.778 \pm 0.007$ GeV as predicted by the chiral unitary approach [28,29]

The particle masses used in this work are calculated based on a theoretical model, namely the chiral unitary approach. These calculations involve model parameters, which directly affect the particle masses. We further evaluate the impact of variations in these parameters on the calculated cross sections. The theoretical uncertainties of the particle masses are derived from Refs. [28,29], and the specific values are presented in Table 1. Within the range of mass variations, we have calculated their production cross sections, and the results are shown in Fig. 5. It is evident that the mass uncertainties have a negligible effect on the production cross sections of $B_{s0}^*$ and $B_{s1}^*$. For a quantitative assessment, we examine the cross sections at an incident kaon momentum of 14.0 GeV. For the $B_{s0}^*(5725)$ production channel, the mass uncertainty $\Delta M = \pm 0.039$ GeV results in cross sec-tions ranging from approximately 0.0327 nb to 0.0332 nb for $\alpha = 2.90$. Due to the significantly tighter mass constraint of $B_{s1}^*$ ($\Delta M = \pm 0.007$ GeV), its mass uncertainty has a much smaller effect on the cross sections. Specifically, the cross sections range from 0.0884 to 0.0891 nb for $\alpha = 2.90$.

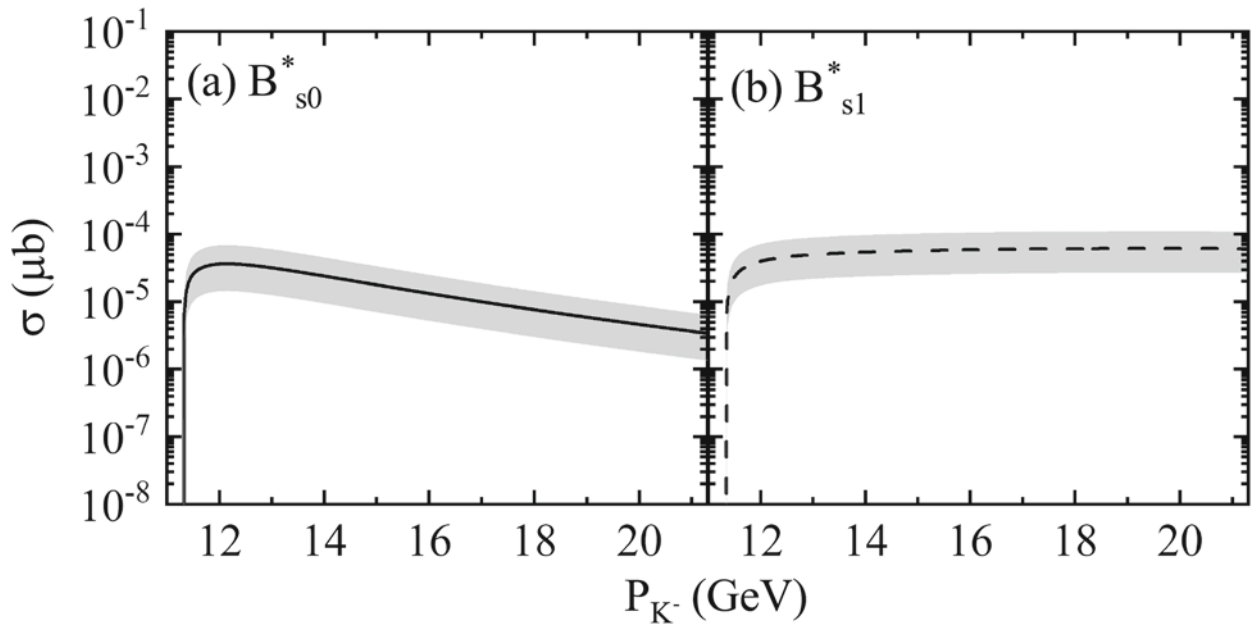

**Fig. 6** The total cross sections for the processes **a** $K^-p \to \Lambda_b^0 B_{s0}^*$ and **b** $K^-p \to \Lambda_b^0 B_{s1}^*$ are evaluated for $\alpha = 2.90$. The gray bands represent the uncertainty in the cross sections arising from the theoretical uncertainties in the coupling constants $g_{B_{s0}^* B\bar{K}}$ and $g_{B_{s1}^* B^*\bar{K}}$. The data has been organized in Table 1

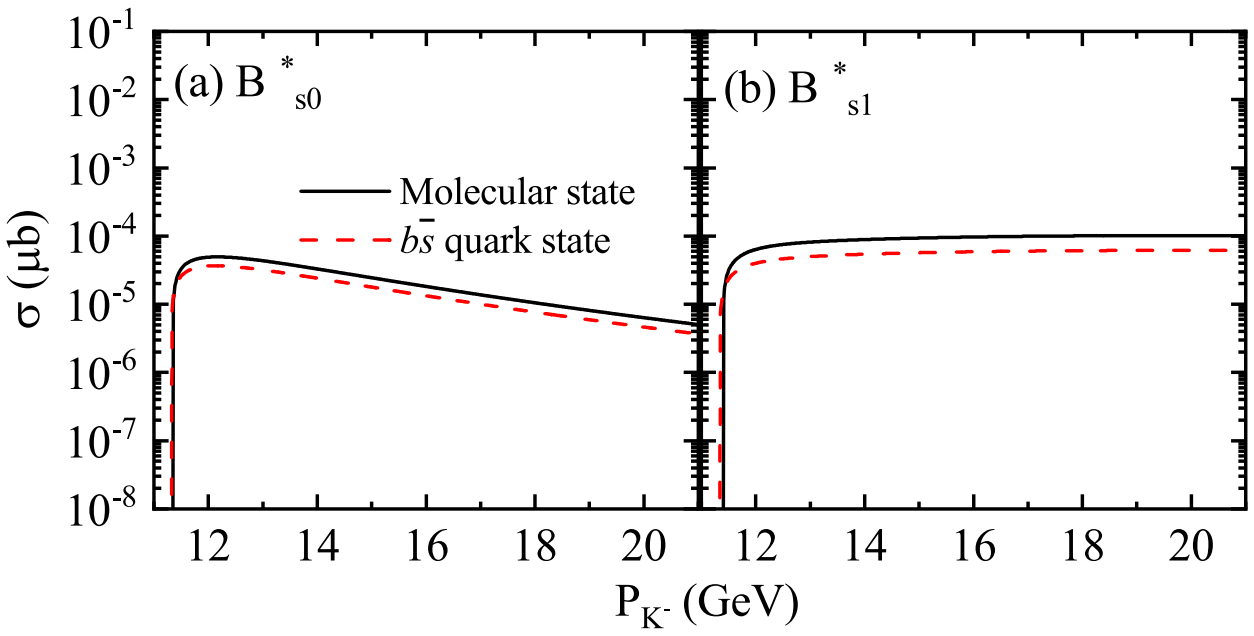

**Fig. 7** The total cross sections for the production of $B_{s0}^*$ (**a**) and $B_{s1}^*$ (**b**) in $K^-p$ collisions with $\alpha = 2.90$, comparing the predicted yields for the hadronic molecular (solid) and conventional $b\bar{s}$ (dashed) assignments

In order to unambiguously identify the experimentally observed structures as the $B\bar{K}$ and $B^*\bar{K}$ molecular states-corresponding to the yet-unobserved $B_{s0}^*$ and $B_{s1}^*$, respectiv ely-it is essential to rule out alternative explanations. In particular, conventional excited mesons composed of a $b\bar{s}$ quark pair, namely $B_{s0}^*$ and $B_{s1}^*$, may also exist in this energy region. Indeed, Refs. [41,42] predicted the existence of two $b\bar{s}$ mesons with masses $M_{B_{s0}^*} = 5700$ MeV and $M_{B_{s1}^*} = 5720$ MeV using QCD sum rules. Therefore, a careful distinction between the properties of these conventional states and those of the molecular configurations is necessary. In this work, we compare the production cross sections of these states in the reactions $K^-p \to \Lambda_b^0 B_{s0}^*$ and $K^-p \to \Lambda_b^0 B_{s1}^*$ under both the compact $b\bar{s}$ configuration and the hadronic molecular hypothesis, thereby providing theoretical criteria for distinguishing their internal structures in future experiments.

Using the results from Ref. [42], the coupling constants of these conventional quark states to the $B\bar{K}$ and $B^*\bar{K}$ channels were determined to be $g_{B_{s0}^* B\bar{K}} = 20.0 \pm 7.4$ GeV and $g_{B_{s1}^* B^*\bar{K}} = 18.1 \pm 6.1$ GeV, and the data are also presented in Table 1. Based on these values, we calculate the production cross sections for the reactions $K^-p \to \Lambda_b^0 B_{s0}^*(5700)$ and $K^-p \to \Lambda_b^0 B_{s1}^*(5720)$, under the assumption that $B_{s0}^*(5700)$ and $B_{s1}^*(5720)$ are conventional quark–antiquark states. The results are presented in Fig. 6. It is evident that the theoretical uncertainties in the coupling constants $g_{B_{s0}^* B\bar{K}}$ and $g_{B_{s1}^* B^*\bar{K}}$ introduce a substantial variation in the predicted cross sections. To quantify this effect, we examine the total cross section at an incident $K^-$ meson momentum of 14.0 GeV. For $B_{s0}^*$ production with $\alpha = 2.90$, the cross section ranges from 0.00957 to 0.0452 nb due to the coupling constant uncertainties, corresponding to a spread factor of 4.72. A similar behavior is observed for $B_{s1}^*$ production, where the uncertainties in the coupling constants lead to a cross-section variation factor of approximately 4.06.

When comparing the production cross sections of $K^-p \to \Lambda_b^0 B_{s0}^*$ and $K^-p \to \Lambda_b^0 B_{s1}^*$, we find that, using the central values of the masses and coupling constants listed in Table 1, the cross sections for the molecular configurations are slightly larger than those for the conventional quark–antiquark states, as shown in Fig. 7. Specifically, at $P_{K^-} = 14.0$ GeV, the cross section for the molecular $B_{s0}^*(5725)$ is 0.0329 nb, compared to 0.0241 nb for the conventional $B_{s0}^*(5700)$. Similarly, the molecular $B_{s1}^*(5778)$ yields a cross section of 0.0888 nb, whereas the conventional $B_{s1}^*(5720)$ gives 0.0541 nb. This difference remains essentially constant (1.36 for $B_{s0}^*$ and 1.64 for $B_{s1}^*$ production) as the incident kaon energy increases. Therefore, under the present mechanism, it is difficult to draw a definitive conclusion on whether the observed $B_{s0}^*$ and $B_{s1}^*$ states are molecular or conventional quark–antiquark states based solely on the variation of the incident energy. Nevertheless, although our current results cannot unambiguously determine the internal structures of these states, they provide useful guidance for the search of these two as-yet unobserved particle states.

In addition to the total cross sections for the processes $K^-p \to \Lambda_b^0 B_{s0}^*$ and $K^-p \to \Lambda_b^0 B_{s1}^*$, we have also calculated their differential cross sections. Figure 8 shows the differential cross sections ($d\sigma/d\cos\theta$) for these reactions at beam energies $P_{K^-} = 12$, 14, and 16 GeV, with and without initial-state interactions (ISI). Without ISI, the angular distributions are relatively flat, indicating nearly isotropic production dominated by t-channel $B^{(*)}$ exchanges. When ISI are included, the distributions develop a pronounced double-peaked structure, with maxima at forward ($\cos\theta \approx +1$) and backward ($\cos\theta \approx -1$) angles and a minimum near central angles ($\cos\theta \approx 0$). The double-peaked feature becomes more prominent at higher beam energies, reflecting the increasing impact of ISI. Moreover, ISI enhance the overall differential cross sections by roughly an order of magnitude.

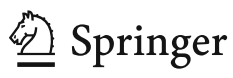

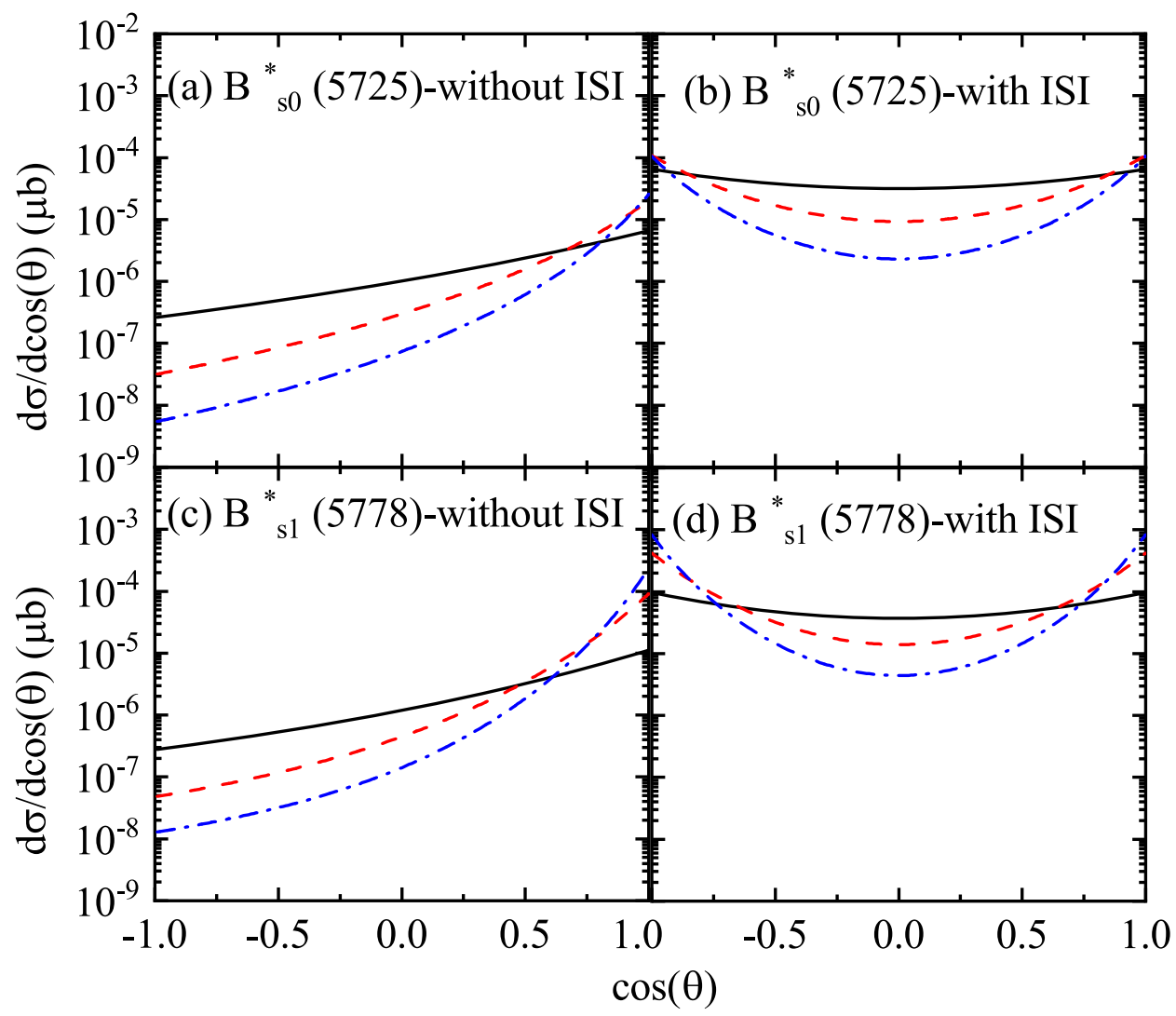

**Fig. 8** Differential cross sections for **a**, **b** $K^-p \to \Lambda_b^0 B_{s0}^*$ and **c**, **d** $K^-p \to \Lambda_b^0 B_{s1}^*$, calculated with $\alpha = 2.90$. The black solid lines, red dashed lines, and blue dashed-dotted lines are obtained at beam energies $P_{K^-} = 12$, 14, and 16 GeV, respectively

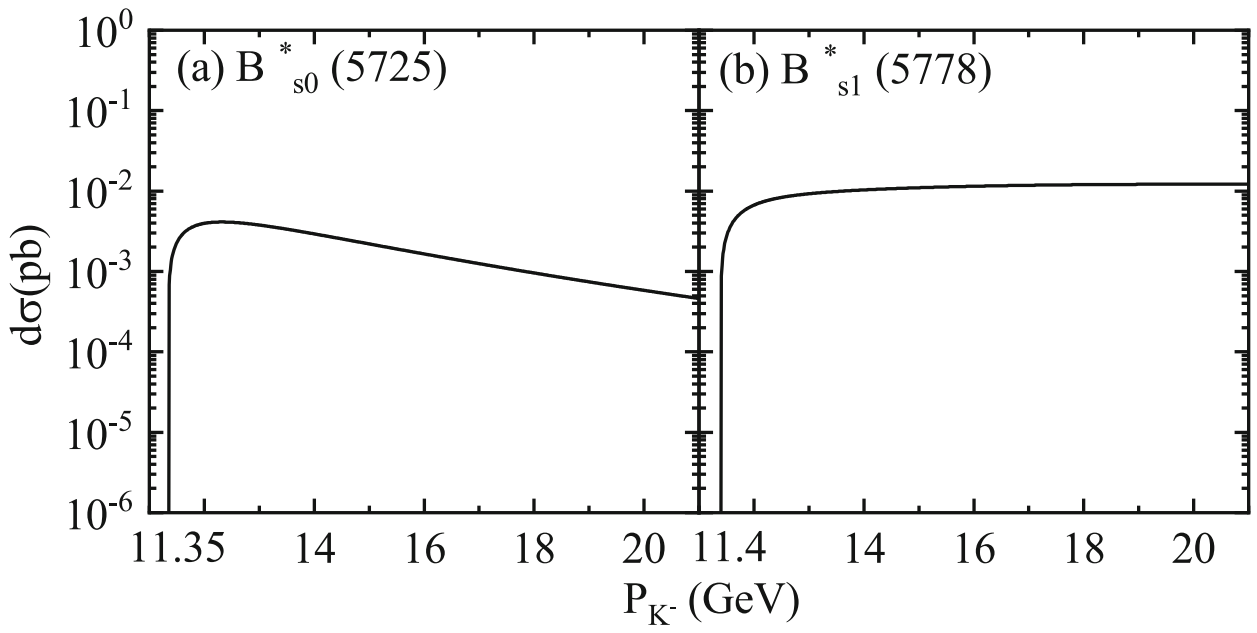

**Fig. 9** The total cross sections for the processes **a** $K^-p \to \Lambda_b^0 B_{s0}^* \to \pi B_s^0 \Lambda_b$ and **b** $K^-p \to \Lambda_b^0 B_{s1}^* \to \pi B_s^{*0} \Lambda_b$ are evaluated for $\alpha = 2.90$

Finally, we suggest searching for the two $B_{s0/s1}^*$ states by reconstructing the invariant mass spectra of the $\pi B_s^{(*)}$ system, which corresponds to their dominant decay channels, in the reaction $K^-p \to B_{s0/s1}^* \Lambda_b^0 \to \pi B_s^{(*)} \Lambda_b^0$. This process has a significant advantage in providing a very clean experimental signal, since at present there are no other known particles that can be produced in $K^-p$ interactions and subsequently decay into the same $\pi B_s^{(*)}$ final states, thereby ensuring that the background contribution is negligible. We estimate the such reaction cross section using the following expression [57,58]

$$\frac{d\sigma_{K^-p\to\Lambda_b^0\pi B_s^{(*)}}}{dM_{\pi B_s^{(*)}}} = \frac{2m_{B_{s0/s1}^*} M_{\pi B_s^{(*)}}}{\pi} \times \frac{\sigma_{K^-p\to\Lambda_b^0 B_{s0/s1}^*} \Gamma_{B_{s0/s1}^* \to \pi B_s^{(*)}}}{(M_{\pi B_s^{(*)}}^2 - m_{B_{s0/s1}^*}^2)^2 + m_{B_{s0/s1}^*}^2 \Gamma_{B_{s0/s1}^*}^2} \tag{17}$$

Based on the theoretically calculated total decay widths $\Gamma_{B_{s0/s1}^*}$ and their partial widths for the $\pi B_s^{(*)}$ decay channels [28,29], we estimate the relevant reaction and show the numerical results in Fig. 9. We find that their cross sections are of the order of 0.01 pb, which makes them potentially detectable in experiments.

## 4 Summary

We investigate the production mechanisms of the resonances $B_{s0}^*$ and $B_{s1}^*$ in the reactions $K^-p \to \Lambda_b^0 B_{s0}^*$ and $K^-p \to \Lambda_b^0 B_{s1}^*$ using an effective Lagrangian approach. These processes are primarily driven by the $t$-channel exchange of $B$ and $B^*$ mesons. We first consider the scenario in which these states are heavy-quark flavor partners of the well-known $D_{s0}(2317)$ and $D_{s1}(2460)$, which have not yet been observed but remain a focus of experimental studies. The coupling constants $B_{s0}^* \to B\bar{K}$ and $B_{s1}^* \to B^*\bar{K}$ are taken from chiral unitary theory [28,29]. Our results indicate that their production cross sections can reach approximately 0.01 nb, which is within the reach of current and future experiments such as AMBER@CERN and J-PARC. Furthermore, we suggest that $B_{s0}^*$ be measured at an incident $K^-$ beam momentum of around 12.18 GeV, while higher beam energies are more suitable for detecting $B_{s1}^*(5778)$.

We also calculate the production cross sections of $B_{s0}^*$ and $B_{s1}^*$ as conventional $b\bar{s}$ mesons. These values are slightly lower than those in the molecular scenario but still within the observable range. However, this difference is not pronounced, and thus our results alone cannot determine whether the observed $B_{s0/s1}^*$ states are molecular or conventional quark states.

In our calculations, we find that the initial-state interaction (ISI) effects, realized via Pomeron and Reggeon exchanges, are very important, as they can enhance the cross section by approximately one order of magnitude. At the same time, they lead to significant changes in the angular distribution of the differential cross section, revealing a pronounced double-peaked structure due to ISI effects, which is absent in the angular distributions calculated without ISI. These findings provide a unique signal for the search of the currently widely discussed but yet unobserved $B_{s0}^*$ and $B_{s1}^*$ states.

**Acknowledgements** This work was not supported by any funding agency.

**Data Availability Statement** Data will be made available on reasonable request. [Author's comment: The datasets generated during and/or analysed during the current study are available from the corresponding author on reasonable request.]

**Code Availability Statement** Code/software will be made available on reasonable request. [Author's comment: The code/software generated during and/or analysed during the current study is available from the corresponding author on reasonable request.]

Funded by SCOAP[3].